\documentclass{jpsj-suppl}
\usepackage{txfonts} 
\usepackage[export]{adjustbox}

\newcommand{\apj}{ApJ}
\newcommand{\prl}{Phys. Rev. Lett.}
\newcommand{\apjs}{ApJS}
\newcommand{\nature}{Nature}
\newcommand{\aanda}{A\&A}

\title{Effects of Dimensionality on Pair-Instability Supernova Explosions}

\author{Matthew S. \textsc{Gilmer}$^{1}$, Alexandra \textsc{Kozyreva}$^{2}$, Raphael \textsc{Hirschi}$^{2,3,4}$, and Carla \textsc{Fr\"ohlich}$^{1}$}

\inst{$^{1}$Department of Physics, North Carolina State University, Raleigh, NC, 29695-8202, USA \\
$^{2}$Astrophysics Group, Keele University, Keele ST5 5BG, UK \\
$^{3}$Kavli IPMU (WPI),University of Tokyo, 5-1-5 Kashiwanoha, Kashiwa, 277-8583, Japan \\
$^{4}$UK Network for Bridging Disciplines of Galactic Chemical Evolution (BRIDGCE)}

\email{msgilmer@ncsu.edu}

\recdate{August 2 2016}

\abst{Since the emergence of the new class of extremely bright transients, super-luminous supernovae (SLSNe), three main 
mechanisms to power their light curves (LCs) have been discussed. They are the spin-down of a magnetar,  interaction with circumstellar material, and the decay of large amounts of radioactive nickel 
in pair-instability supernovae (PISNe). 
 Given the high degree of diversity seen within the class, it is 
 possible that all three mechanisms are at work. 
PISN models can be self-consistently simulated from the main sequence phase of very massive stars (VMS) through to their explosion. These models robustly predict large amounts of radioactive nickel and thus very luminous SN events. However, PISN model LCs evolve more slowly than even the slowest evolving SLSNe. 
 Multidimensional effects on the ejecta structure, specifically the mixing of radioactive nickel out to large radii, could alleviate this discrepancy with observation. Here we explore the multidimensional effects on the LC evolution by simulating the explosion phase in 1D, 2D, and 3D. We find that the 
ejecta from the multidimensional simulations have slightly shallower abundance gradients due to mixing at shell boundaries. We compute synthetic LCs whose shapes show no discernible differences due to the multidimensional effects.  
 }

\kword{supernovae: general, supernovae: individual (PTF12dam), stars:evolution, stars:massive, hydrodynamics, radiative transfer \ldots}

\begin{document}
\maketitle

\section{Introduction}

A PISN is the explosive death of a VMS with a carbon-oxygen (CO) core in the mass range 64~M$_{\odot}$ $<$ M$_{\mathrm{CO}}$ $<$ 133~M$_{\odot}$ \cite{hw02}. In PISNe, core collapse is triggered by the production of electron-positron pairs from photon collisions. The collapse of the core is subsequently reversed by the explosive burning of oxygen \cite{barkat}. PISNe are rare events that can produce large amounts of radioactive nickel, the decay of which can power extremely bright LCs. For these reasons, PISNe were proposed as explanations for some SLSNe like SN2007bi \cite{gal-yam}. However, PISN models have heretofore been unable to reproduce the fast rise to peak luminosity seen in SLSNe \cite{kasen,chatz}. Here we investigate whether or not a more extended nickel distribution in the ejecta (arising naturally out of multidimensional hydrodynamical simulations) can sufficiently decrease the rise time of PISN model light curves to match that of PTF12dam, one of the more slowly evolving (and well-observed) SLSNe.


\section{Methods}

PISN simulations were carried out with FLASH (version 4.3) \cite{flash} in 1D, 2D, and 3D for two VMS models (P200 and P250). The VMS models used here were computed with the GENEC stellar evolution code \cite{genec}. Both models are non-rotating models and have initial metallicity $Z=0.001$. Models P200 and P250 began their lives as 200~M$_{\odot}$ and 250~M$_{\odot}$ stars. Model P200 lost almost all of its hydrogen envelope and ended its life as a 100.9~M$_{\odot}$ helium core while model P250 lost its entire hydrogen envelope, along with almost all of its helium envelope, and ended its life as a 126.7~M$_{\odot}$ compact CO core. These models are also the main models in a forthcoming paper about their applicability to PTF12dam \cite{sasha}. FLASH followed the core collapse, explosion, and expansion phases right up until the moment before shock breakout (SBO). At this point the 1D and 3D (angular-averaged) data, including the entire ejecta, were mapped into the 1D radiation-hydrodynamics code STELLA \cite{stella} and synthetic LCs were computed.

\section{Results}

Table \ref{tab:explosion_properties} shows the explosion properties from the \verb|FLASH| simulations for both models. The explosion strengths, as evidenced by the nickel yields and explosion energies, were weaker for higher dimensionality. This is a consequence of an under-resolved computational grid for the multidimensional simulations. We expect the multidimensional explosion properties to converge to values similar to those of the 1D case at some higher resolution. More details will be given in the forthcoming paper \cite{me}.

\begin{table}[tbh]
\centering
\caption{Explosion properties for model P200 and P250 are shown for 1D, 2D, and 3D runs.}
\label{tab:explosion_properties}
\begin{tabular}{l|llllllll}
\hline
   & \multicolumn{2}{l}{Collapse Time (s)} & \multicolumn{2}{l}{Silicon Yield (M$_{\odot}$)} & \multicolumn{2}{l}{Nickel Yield (M$_{\odot}$)} & \multicolumn{2}{l}{Explosion Energy (B)} \\
   & P200 & P250 & P200 & P250 & P200 & P250 & P200 & P250 \\
\hline
1D & 24.2 & 17.6 & 22.0 & 24.1 & 15.6 & 36.0 & 53.8 & 84.9 \\
2D & 24.9 & 17.4 & 22.4 & 24.7 & 10.2 & 33.2 & 48.8 & 82.6 \\
3D & 25.0 & 17.4 & 22.4 & 24.9 &  8.5 & 32.3 & 47.1 & 81.7 \\
\hline
\end{tabular}
\end{table}

\begin{figure}[tbh]
\centering
\begin{tabular}{ll}
\hspace{-.015\textwidth}\includegraphics[width=0.48\textwidth]{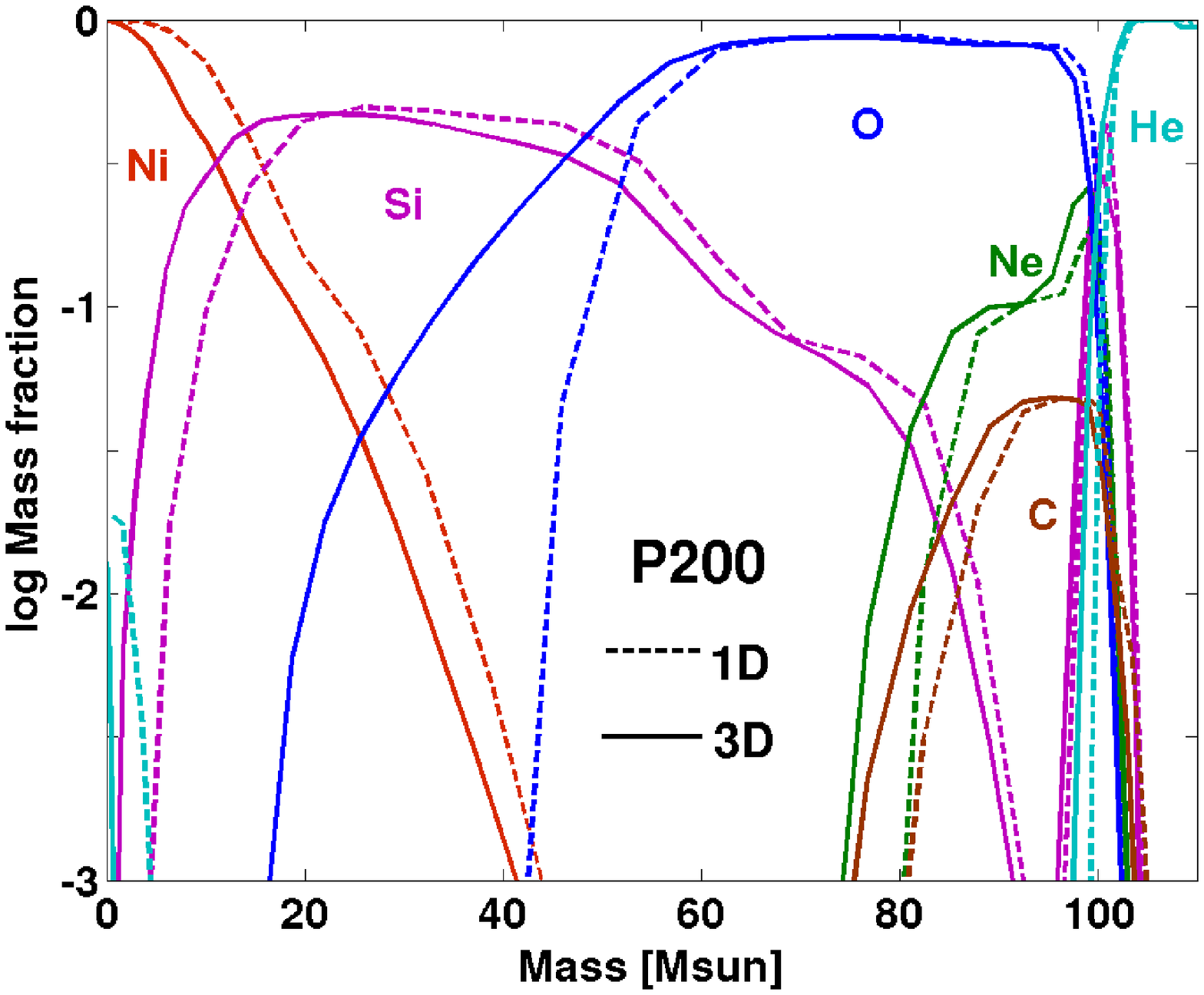} & \includegraphics[width=0.48\textwidth]{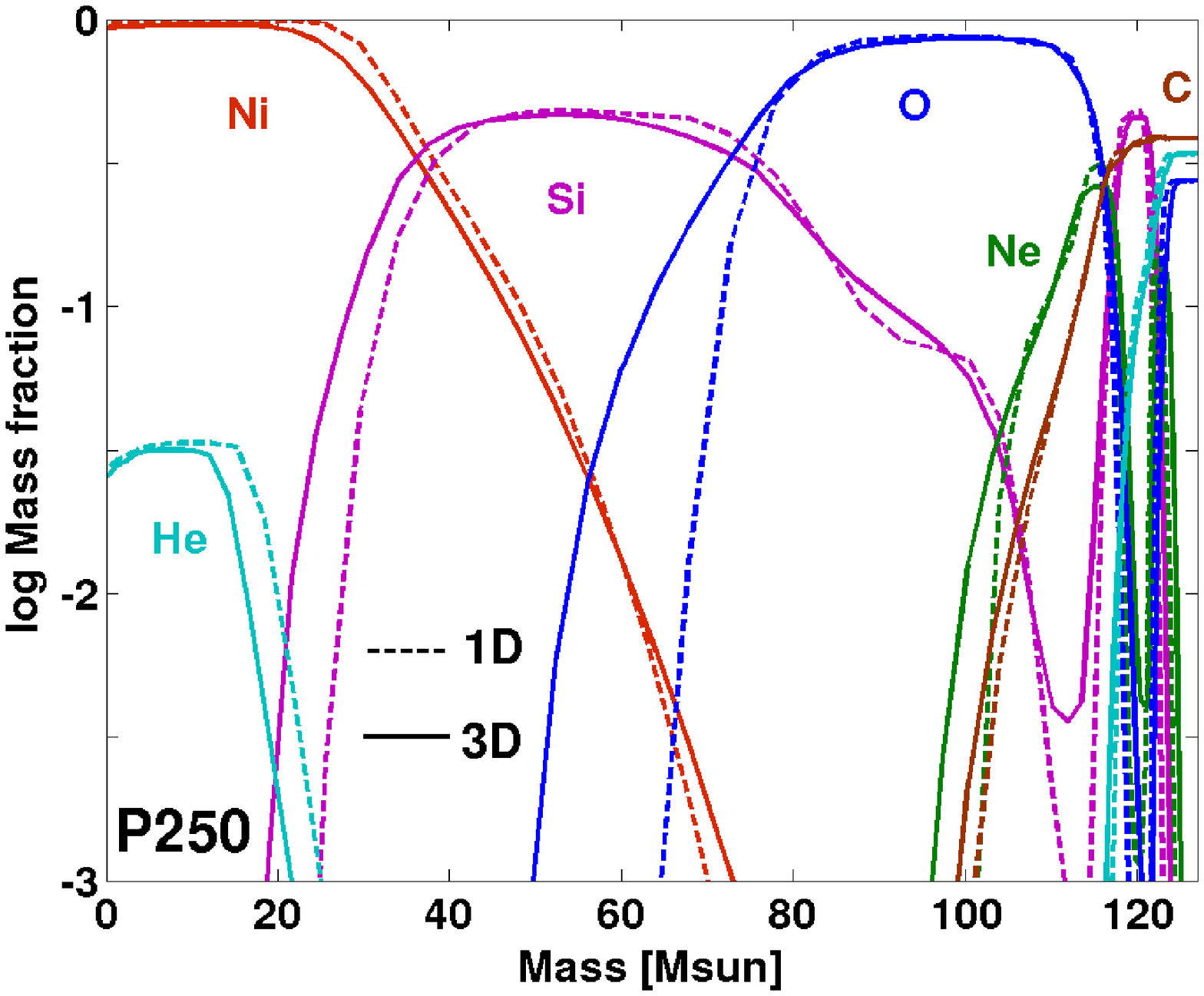} \\
\end{tabular}
\caption{The 1D (dashed) and 3D (solid) pre-SBO FLASH abundance profiles for models P200 (left panel) and P250 (right panel) at the moment before mapping into STELLA.}
\label{f1}
\end{figure}

The abundance profiles were slightly altered by simulating in multiple dimensions due to the formation of Rayleigh-Taylor instabilities at the interfaces between shells of differing composition. The 1D and 3D pre-SBO abundance profiles for both models are shown in Figure \ref{f1}. In general, the 3D profiles exhibit shallower abundance gradients, especially at the Si-O interfaces. Mixing at this interface should not affect the LC shape, though it could alter the spectrum (specifically the shape of the neutral oxygen emission lines). We expect mixing at the Ni-Si interface to have the strongest effect on the LC shape since this would lead to a more extended nickel distribution. The effect is most easily seen for model P250 in the right panel of Figure \ref{f1} where the 3D nickel distribution extends further in mass coordinate (note the log scale) despite synthesizing less nickel overall.

\begin{figure}[tbh]
\centering
\includegraphics[width=0.6\textwidth]{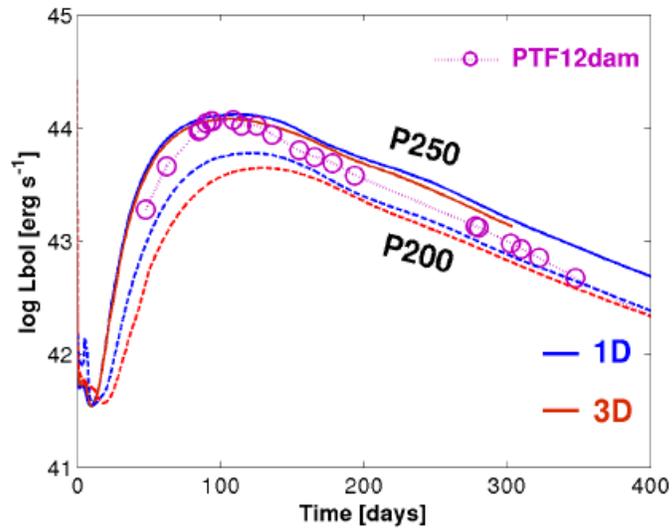}
\caption{The 1D (blue) and 3D (red) STELLA LCs for models P200 (dashed) and P250 (solid) are compared to that of PTF12dam.}
\label{f2}
\end{figure}

The synthetic LCs computed from the 1D and 3D (angular-averaged) \verb|FLASH| profiles for both models are plotted against observations of PTF12dam in Figure \ref{f2}. The differences between the 1D and 3D LC shapes, for both models, can be explained by the differences in the nickel yields (lower in the 3D case) while any multidimensional effect on the rise times is indiscernible. Further study, requiring computational resources with greater memory, is needed to determine whether or not this result will hold for higher resolution multidimensional simulations.

\section*{Acknowledgments}

MG and CF are supported through an Early Career Award by the US Department of Energy (grant no. SC0010263). The authors acknowledge support from EU-FP7-ERC-2012-St Grant 306901. RH acknowledges support from the World Premier International Research Center Initiative (WPI Initiative), MEXT, Japan.


\end{document}